# Effect of impurity on the determination of ground-state properties of parabolic quantum dot composed of N electrons

Berna GÜLVEREN[1], Ülfet ATAV[1], Mehmet SAHIN[1], Mehmet TOMAK[2]

[1]Selçuk University, Faculty of Arts and Science, Physics Dept. Campus 42075 Konya, Turkey
[2]Middle East Technical University, Physics Dept. 06531 Ankara, Turkey

The two dimensional Thomas-Fermi approximation is applied to the problem of parabolic quantum dot composed of N electrons and an on-center impurity. Change induced by impurity on electron density, chemical potential and total energy is discussed and it is found that existence of impurity makes significant changes on the determination of above properties especially for the small number of electrons and strong confinements.

## I- INTRODUCTION

In recent years, progress in nanotechnology has allowed manufacturing of ultra small systems in the areas of electronics and optoelectronics with a high precision capability of controlling the size and shape and hence the number of electrons (N) confined in these ultra small systems. Quantum dots are the typical examples of these systems where the electrons are confined in all three dimensions. Confinement can be caused by a difference in the material composition or by an externally applied biasing potential, or both[1-2]. Because of their finite size, quantum dots may contain finite number of electrons. When the number of electrons is small, the electronic properties of the system depend strongly on N and quantum mechanical effects become rather important. On the other hand, as the number of particles is increased, the system acts as a confined electron cloud ($N \rightarrow \infty$) and the methods of statistical mechanics become applicable. Thomas-Fermi model and it's modifications has been used by several authors [3-8] for evaluation of electronic and other properties of these complicated systems as well as of various other many fermion systems, e.g., in the treatment of atoms, molecules, and nuclei.

Evaluation of the properties of a quantum dot requires a self consistent solution of the Schrödinger equation for an electron gas in two dimensions. Bhaduri et.al.[9] solved exactly the two dimensional equation in conjunction with Poisson equation for atoms and ions. Pino has obtained an analytical solution of the Thomas-Fermi equation for the two dimensional parabolic dot where the electrons interact through a logarithmic potential in the absence and presence of magnetic field[10-11].

Collective excitations in a quantum dot containing finite number of electrons were investigated by Sinha[12] for three different types of electron-electron interaction within the Thomas-Fermi approximation. Some properties like total energy, chemical potential and differential capacitance were calculated for three dimensional quantum dot arrays using soft model confinement under the local Thomas-Fermi-Dirac assumption[13]. Moreover, Thomas-Fermi approach had been applied to the problem of two dimensional devices including Dirac's local exchange and Fermi-Amaldi's exchange correction and numerical examples were given for arrays of quantum dots using Gaussian confining potential[14]. A semiclassical approach to the ground state and density oscillation of two dimensional nanostructures of arbitrary shape was discussed by Puente et al. including a Weizsacker gradient term[15]. For the arbitrary effective interactions, it was shown that Thomas-Fermi method provides a good framework for studying the properties of many electrons confined in a quantum dot and corrections to the method produce some shell effect like changes on Thomas-Fermi energy[16].

Many theoretical and experimental studies of impurity related properties in low dimensional heterostructures have been reported in the last decade[17-34]. The great interest for understanding of the properties of these impurity containing systems come from that the impurity modify the energy levels of materials and in turn affect their electronic and optical properties. So these systems have potential use in designing electro-optical devices. Porras et.al.[17-18] have calculated impurity binding energies in spherical GaAs - GaAlAs quantum dots as a function of dot radius using variational procedure within the effective-mass approximation. By using the method of series expansion, Zhu et.al.[19-20] calculated the binding energy of an impurity at the center of a quantum dot for infinite potential. Ground-state impurity binding energy and density of impurity states in a cubic GaAs quantum dot have been calculated by Ribeiro and Latge [21]. In another study, Yang et.al.[22] calculated the fine structure of the energy levels for a hydrogenic impurity located at the center of a quantum dot using a simpler exact solution for the potential well. The problem of a donor impurity in a confined geometry with dielectric mismatch at the boundaries has been studied by Ferreyra and Proetto[23]. The influence of the hydrogenic impurity potential and the role of electron-electron interactions in the formation of electron structure in the quantum dot have been investigated by Lee et.al.[24]. Very recently, a numerical method is proposed by Yau and Lee to calculate the low-lying energy spectra of N-electron polar quantum dot bound to an on-center Coulomb impurity [25].

In this study we consider the interesting problem of understanding how an impurity located at the center of a parabolic two dimensional quantum dot affects the chemical potential, density distribution and the total energy of electrons confined in the dot. We make use of two dimensional Thomas-Fermi approximation. In section II, we give the analytical solution of the coupled Thomas-Fermi equation and Poisson's equation in cylindricall coordinates, then we obtain the physical properties such as electron density, dot radius, chemical potential and total energy of the system . In section III, we compare our results with the ones that were calculated by Pino[10] when there is no impurity. We expand our discussion to see how the number of particles and the strength of confinement affect the behavior of above properties. In all calculations, we use atomic units where $\hbar = e = m_e = 1$.

## II. THEORY

In the effective mass approximation, two dimensional Thomas-Fermi equation for an electron gas and on-center impurity confined in a quantum dot can be written as

$$\frac{p_f^2}{2m^*} + \upsilon + V = \mu, \tag{1}$$

where $m^*$ is the effective mass of the electron, $p_f$ is the Fermi momentum, $\upsilon$ is the external (confining) potential, $\mu$ is the chemical potential and $V$ is the electrostatic potential given by

$$V(r) = \frac{1}{\varepsilon} \ln\left(\frac{r}{a}\right) - \frac{1}{\varepsilon} \int n(r') \ln\frac{|r-r'|}{a} d^2r'. \tag{2}$$

Here, $\varepsilon$ is the dielectric constant of the material and $a$ is an arbitrary constant corresponding to a radius at which the potential vanishes. The form of the potential used in Eq.(2) has been chosen so that it is consistent with the studies of two dimensional Coulomb gas problem[35] and it satisfies the Poisson equation:

$$\nabla^2 V = -\frac{2\pi}{\varepsilon} n(r), \tag{3}$$

where n(r) is the electron density.

Within the Thomas-Fermi approach, Fermi momentum can be written in terms of n(r) as

$$p_f = (2\pi n(r))^{1/2}, \tag{4}$$

for spin −1/2 particles. Then combining Eqs.(1) and (4), n(r) can be written as

$$n(r) = \left(\frac{m^*}{\pi}\right)[\mu - V(r) - \upsilon(r)], \qquad r \leq r_0$$
$$= 0, \qquad r > r_0 \tag{5}$$

$$N = \int n(r) d^2r, \tag{6}$$

where $r_0$ is the classical turning point at which the electron density vanishes. For convenience we have chosen the constant $a$ of Eq. (2), to be equal to $r_0$, then the equation for $r_0$ becomes

$$\upsilon(r_0) = \mu. \tag{7}$$

When the confining potential is of the parabolic form ( i.e. $\upsilon(r) = \frac{1}{2} m^* \omega^2 r^2$ with $\omega$ being the strength parameter of the potential ), the following equation for $z(x) = n(x) - \frac{m^* \omega^2 \varepsilon}{\pi}$ can be obtained by substituting Eq.(5) in to Eq. (3) and employing a change of variables:

$$\frac{d^2 z}{dx^2} + \frac{1}{x}\frac{dz}{dx} - \xi^2 z(x) = 0. \tag{8}$$

Here, $x = \frac{r}{r_0}$ and $\xi^2 = \frac{2 m^* r_0^2}{\varepsilon}$. The solution of Eq.(8) for $n(x)$ in terms of modified Bessel functions [36-37] can be given as

$$n(x) = A I_0(\xi x) + B K_0(\xi x) + \left(\frac{m^* \omega^2 \varepsilon}{\pi}\right), \qquad x \leq 1$$
$$n(x) = 0. \qquad x \rangle 1 \tag{9}$$

Here, $A$ and $B$ are constants, and can be determined by examining the behavior of the functions $I_\nu(x)$ and $K_\nu(x)$ at the limiting values of $x$ [36].

In the limit $x \langle\langle 1$, $I_\nu(x)$ behaves like the $\gamma$'th power of x and $K_\nu(x)$ increases logarithmically i.e.,

$$I_\nu(x) \approx \left(\frac{1}{2}x\right)^\nu \bigg/ \Gamma(\nu+1), \, (\nu \neq -1, -2, ....)$$

$$K_\nu(x) \approx -\left\{\ln(\frac{1}{2}x) + \gamma\right\},$$

$$\gamma = \text{Euler's constant} = 0.5772.... \quad . \tag{10}$$

The last condition implies that, the electron density at the center of the dot diverges to infinity logarithmically. However, the physically meaningful quantity is the probable number of electrons contained in a volume element *dV* of the space i.e. *n(x) dV*. In our two dimensional case *dV* is proportional *to x dx*, then the probable number of electrons between *x* and *x+dx* is proportional to *n(x) x dx* and the function *x n(x)* is finite for all values of *x* even when the constant B does not vanish. So we will keep both of the constants A and B in all subsequent calculations.

Examining the electrostatic potential as $x \to 0$ also convenient for us at this step to find out A and B. From Eqs.(2), (5) and (9), following equality is obtained

$$\mu - \frac{\pi}{m^*} n(0) = \frac{1}{\varepsilon} \ln(x)\Big|_{x \to 0} - \frac{2\pi r_0^2}{\varepsilon} \int x' \, n(x') \, \ln(x') \, dx'. \tag{11}$$

Without any calculation, constant B can be directly obtained by eliminating the divergent terms from each side of the equation. Then we get B as

$$B = \frac{m^*}{\varepsilon \pi} . \tag{12}$$

The continuity condition at x =1 for n(x) determines A as

$$A = -\left(\frac{m^* \omega^2 \varepsilon}{\pi}\right)\left[1 + \frac{K_0(\xi)}{\varepsilon^2 \omega^2}\right]\frac{1}{I_0(\xi)} . \tag{13}$$

Now using Eqs.(12) and (13), we can rewrite Eq.(9) as

$$n(x) = -\left(\frac{m^* \omega^2 \varepsilon}{\pi}\right)\left[\left(1 + \frac{K_0(x)}{\varepsilon^2 \omega^2}\right)\frac{1}{I_0(\xi)} I_0(\xi x) - \frac{K_0(\xi x)}{\varepsilon^2 \omega^2} - 1\right]. \tag{14}$$

For different numbers of confined electrons N and for different values of confinement strengths, $\xi$ values are determined from the normalization condition

$$\int_0^1 dx \cdot x \cdot n(x) = \frac{m}{\varepsilon} \frac{N}{\pi \xi^2} ,$$

$$\frac{I_1(\xi)}{I_0(\xi)} + \frac{1}{\xi}\frac{1}{\varepsilon^2\omega^2 I_0(\xi)} = \frac{(1-N)}{\xi\varepsilon^2\omega^2} + \frac{1}{2}\xi. \tag{15}$$

(see, e.g., Fig.2). Finally, the energy density functional describing the ground state of the system is given as

$$E = \int d^2r\tau(r) + \frac{1}{2}m^*\omega^2\int r^2 n(r)d^2r + \frac{1}{\varepsilon}\int \ln(\frac{r}{r_0})n(r)d^2r - \frac{1}{2\varepsilon}\int n(r)n(r')\ln\frac{|r-r'|}{r_0}d^2rd^2r' \tag{16}$$

Here, we note that the kinetic energy density within the two dimensional Thomas-Fermi approach can be given as $\tau(r) = \frac{\pi}{2m^*}n(r)^2$. It should also be noted that if we express n(r) in terms of the potentials $\mu$, $\upsilon(r)$ and V(r) then the electron-electron term is eliminated from the expression of the total energy. Then, combining Eqs.14,15 and 16, after a little algebra, results in the following expression for the total energy of the ground state

$$<E_T> = \frac{1}{\xi}\left\{\frac{\xi^2 aN}{8} + \frac{1}{2}a\left[1+\frac{K_0(\xi)}{a}\right]\left[1-\frac{1}{I_0(\xi)}\right] + \frac{1}{2}\left[-K_0(\xi)+\ln 2 - \gamma - \ln(\xi)\right]\right.$$

$$-\frac{a\xi^2}{8} + \frac{a^2\xi^4}{8}\left(-\frac{1}{I_0(\xi)}\right)\left[1+\frac{K_0(\xi)}{a}\right]\left[\frac{I_3(\xi)}{\xi} + \frac{2}{\xi^2}I_2(\xi)\right]$$

$$\left. + \frac{1}{8}a\xi^4\left[-\frac{K_3(\xi)}{\xi} + \frac{4}{\xi^4} + \frac{2}{\xi^2}K_2(\xi) + \frac{a}{4}\right]\right\}, \tag{17}$$

where $a = \varepsilon^2\omega^2$. For a =1, it reduces to

$$<E_T> = \frac{1}{\xi}\left[\frac{\xi^2 N}{8} - \frac{1}{2}\left[\frac{1+K_0(\xi)}{I_0(\xi)}\right]\left[1+(\frac{1}{4})\xi^3\left[I_1(\xi) - \frac{2I_2(\xi)}{\xi}\right]\right] - \frac{1}{8}\xi^3\left[\frac{2K_2(\xi)}{\xi} + K_1(\xi)\right]\right.$$

$$\left. + \frac{1}{8}\xi^2\left[-1+\frac{\xi^2}{4} - \frac{4\ln(\xi)}{\xi^2}\right] + 1 + \frac{\ln 2}{2} - \frac{\gamma}{2}\right]. \tag{18}$$

## III. RESULTS AND DISCUSSION

In this paper, we analyze a system composed of interacting many electrons confined in a parabolic quantum dot and an on-center impurity. Ground state properties of the system like chemical potential, electron density, total energy are calculated within the two dimensional Thomas-Fermi approximation.

Pino[10] has studied the properties of a parabolic quantum dot containing many electrons but he has not considered existence of an impurity. His results are presented in figs.1 and 2 with the dotted lines. It can be clearly seen from figs.1-2 that our results are in good agreement with those of Pino[10] for regions away from the impurity. This is especially true for the cases when the number of confined electrons is high.

In fig.2 we show the variation of the radius of the quantum dot ($\xi$) with the square root of the number of confined electrons $\sqrt{N}$. The dot radius increases almost linearly with $\sqrt{N}$. Noting that the quantum dot considered in this study is two dimensional, this linear increase indicates an electron density independent of the number of confined electrons and constant over the most part of the dot.

In fig.3(a) the electron density within the dot is shown against the relative radial distance ( i.e. the radial distance divided by the dot radius) for various numbers of electrons while the strength of confinement is kept constant. This figure also supports the above conclusion about the electron density based on the $N^{1/d}$ behavior (where d is the number of the dimensions).

In fig.3 (b) variation of the electron density with the radial distance is shown for various numbers of electrons. Here one can clearly see that as the number of electrons increased, the dot expands while keeping the electron density constant within the dot. One can also conclude from fig.3 (b) that the effect of the impurity is limited to a small region near the center of the dot. One might expect this as the electron distribution near the center of the dot is mostly determined by the impurity, and electrons away from the center are not effected due to the screening of the impurity by the electrons near the center. The range of the effect of the impurity is almost constant with respect to N, so that the increase in the number of electrons does not have a significant effect on the electron distribution near the center.

In fig.4(a) and (b) variation of the electron density within the dot is shown for various strengths of the confining potential. The number of confined electrons is 10 in (a) and 50 in (b)-(c). On a first thought one might expect that the effective range of the impurity to be smaller for a stronger confining potential. However the situation in figs.4(a)-(b). and (c) are vice versa, especially for the case with N=10, $\varepsilon^2\omega^2 = 2$ (in fig.4(a)) the effect of the impurity can be seen all over the dot. This can be explained in the following manner. When the strength of confining potential is increased the electron density increases in the intermediate range, naturally this increase is accompanied by a decrease in the number of electrons in the central part of the dot. This weakens the screening effect, and consequently the effective range of the impurity increases . (b) shows that, the increase in the strength of the confining potential results in an increase in the electron density over the intermediate range( where the electron density is almost constant ) and this in turn results in a decrease in the dot radius (c).

Finally we evaluate the total energy for the system using Eqs.( 17-18 ) and in fig.5 we look at the difference in the total energy per particle between the cases with and without impurity for various potential strengths. This energy difference per particle corresponds to the binding energy of the electrons to the impurity. This electron-impurity binding energy first increases with N reaches to a maximum value and then decreases and reaches to a limiting value for $N \rightarrow \infty$. This limiting value is independent of the confining potential, however the confining potential determines the rate of the convergence. Also the value of the maximum of the binding energy depends on the strength of the confining potential. The stronger the potential is the quicker the convergence and the higher the maximum of the binding energy. A similar behavior of the binding energy is observed in a study of hydrogenic impurities in a GaAs - GaAlAs quantum dot[19].

In conclusion, we have determined the physical properties of a parabolic two dimensional quantum dot with an on-center impurity. We have used the two dimensional Thomas-Fermi approach to obtain the electron density and then we obtained other physical properties of the system based on this density. The applicability of the Thomas-Fermi approach is questionable when the number of electrons is small. However, it is encouraging that our results for the binding energy *qualitatively* agree with the results of another study[19] employing a different approach even in the few electron regime.

**FIGURE CAPTIONS**

**FIG.1** Electron density n(x) of a parabolic two dimensional quantum dot for N = 10, $\varepsilon^2\omega^2 = 1$. The unit of n(x) is $\left(\frac{m^*}{\pi\varepsilon}\right)$ and $x = \frac{r}{r_0}$.

**FIG.2** Variation of $\xi$ with respect to N ($\varepsilon^2\omega^2 = 1$).

**FIG.3** Comparison of the density profile of the systems composed of different number of particles, ($\varepsilon^2\omega^2 = 1$). (a) Variation of electron density n(x) with x for N = 10, 50, 100, 150. (b) Variation of electron density n(r) with r for N= 50, 100, 150. The unit of r is $\left(\frac{\varepsilon}{m^*}\right)^{1/2}$.

**FIG.4** Variation of electron density in the radial direction for different strength of confinements. (a) Variation of electron density n(x) with x for $\varepsilon^2\omega^2 = 0.1, 0.5, 1, 2$ and N= 10. (b) Variation of electron density n(x) with x for $\varepsilon^2\omega^2 = 0.5, 1, 2$ and N= 50. (c) Variation of electron density n(r) with the radius of the dot for $\varepsilon^2\omega^2 = 0.5, 1, 2$ and N= 50.

**FIG.5** Change in the total energy per particle induced by impurity $\left(\frac{\Delta E}{N} = \frac{E_{T(with)} - E_{T(without)}}{N}\right)$ for different strength of confinements ($\varepsilon^2\omega^2 = 0.1, 0.5, 1.0, 2.0, 4.0$). The unit of $E_T$ is $\left(\frac{1}{\varepsilon}\right)$.

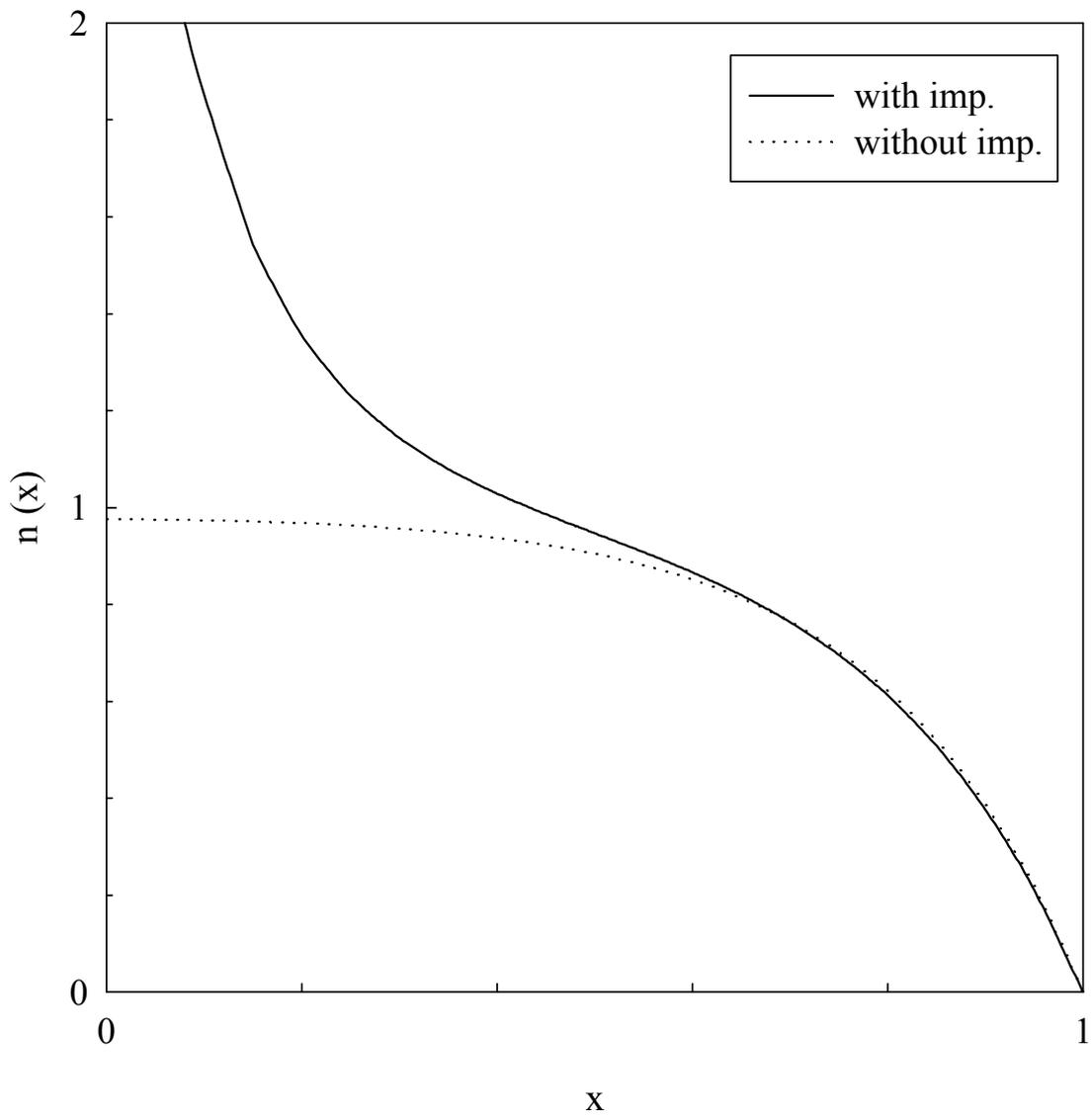

**FIG.1**

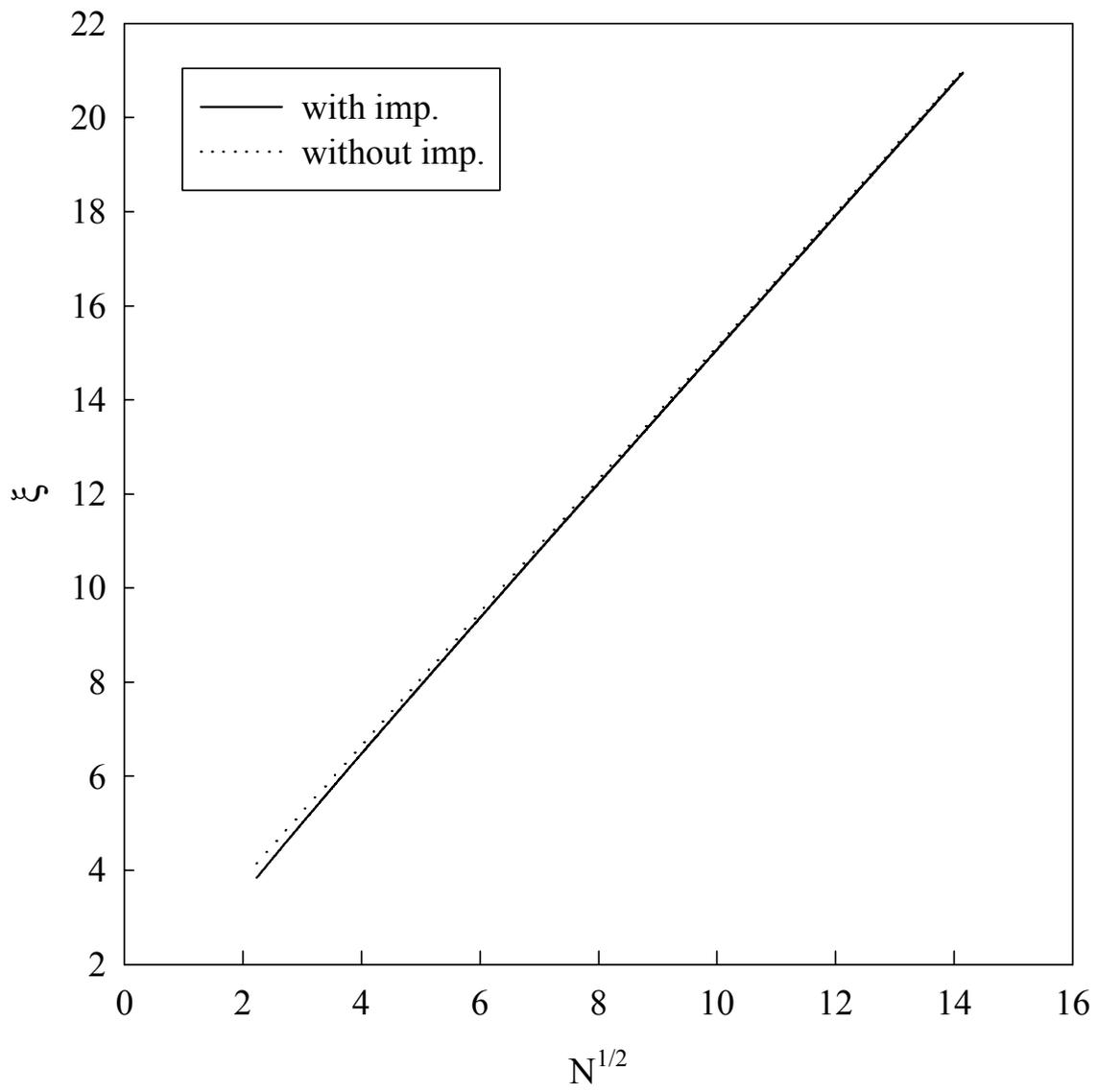

**FIG.2**

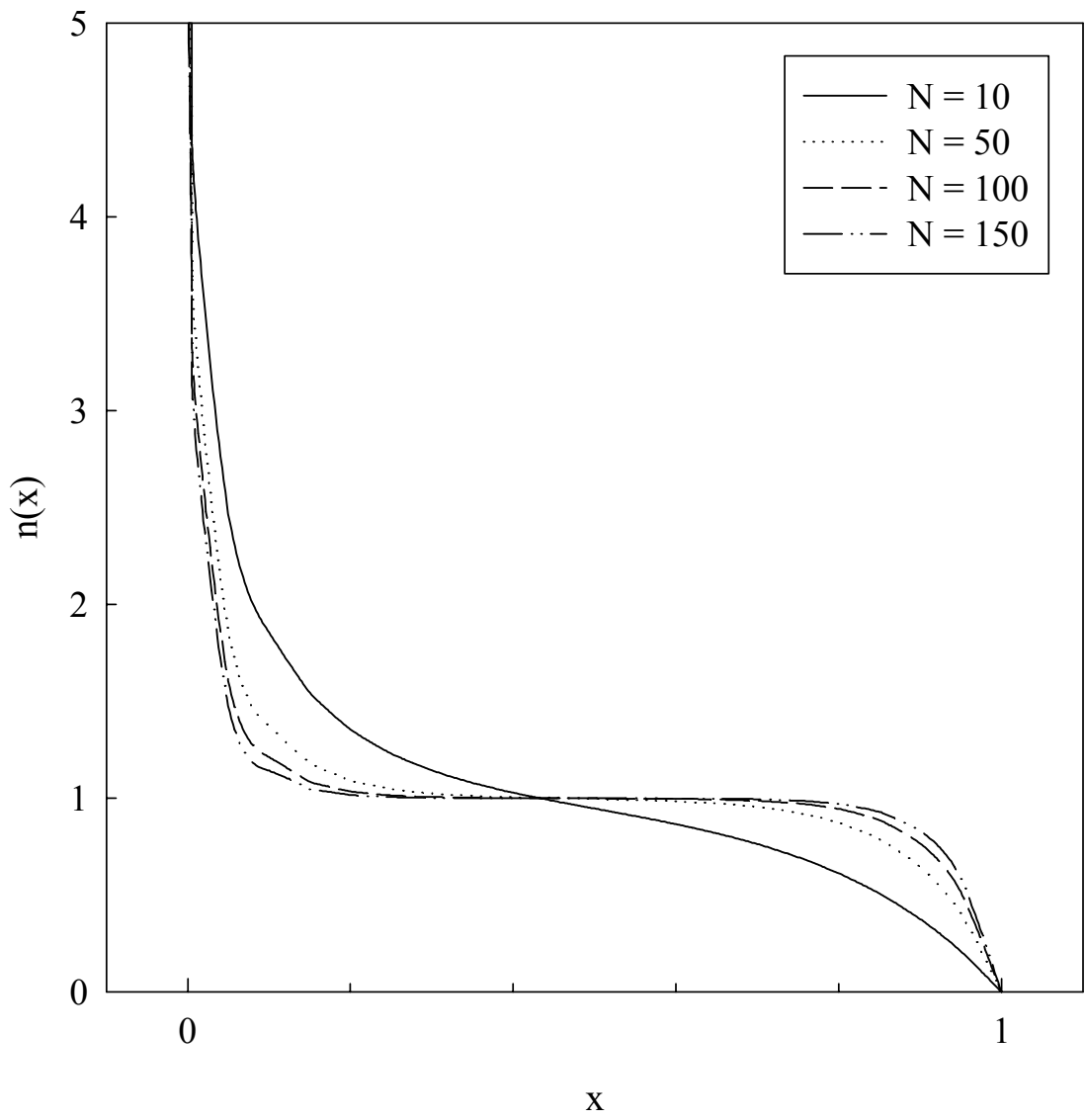

**FIG.3(a)**

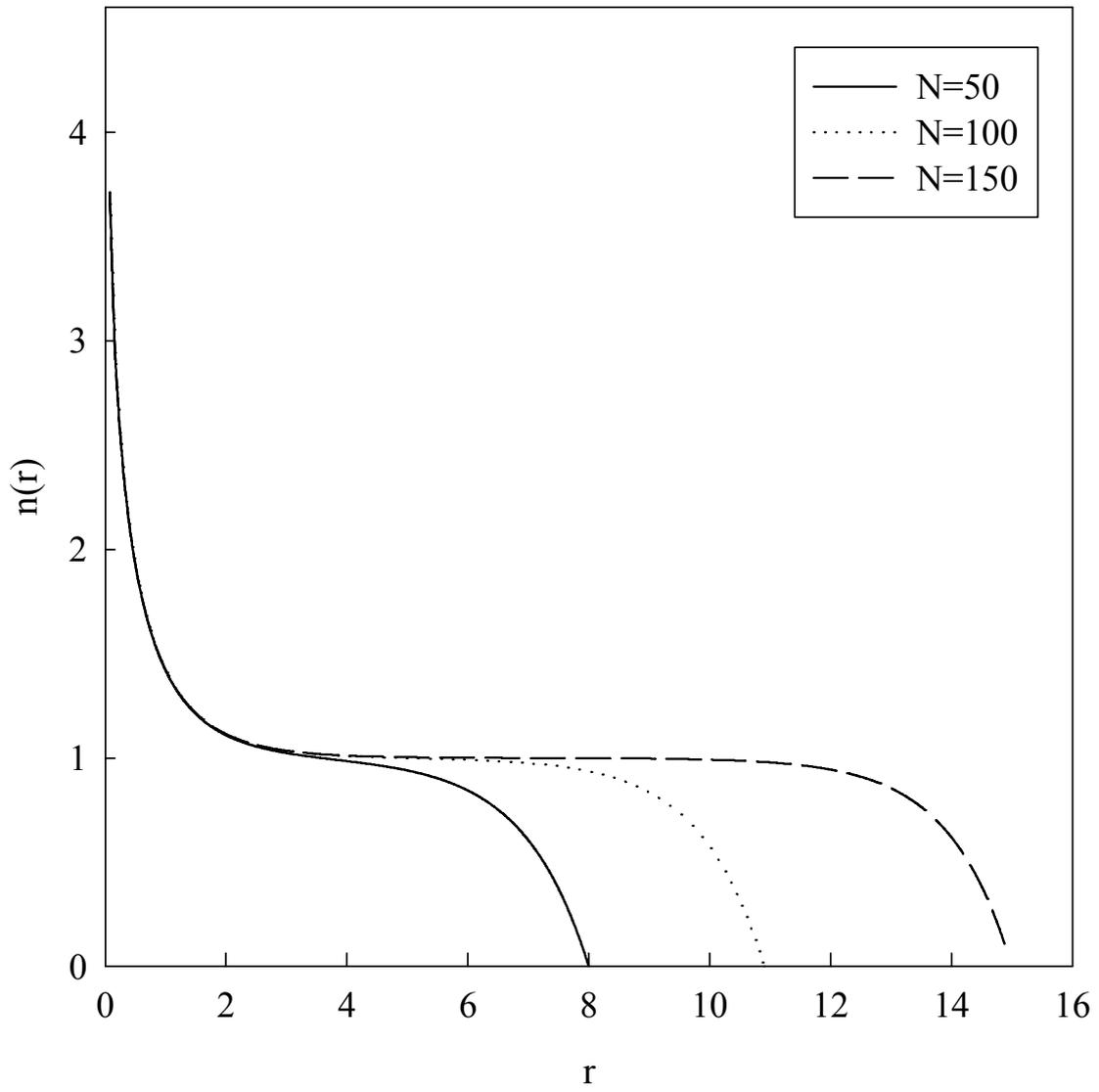

**FIG.3(b)**

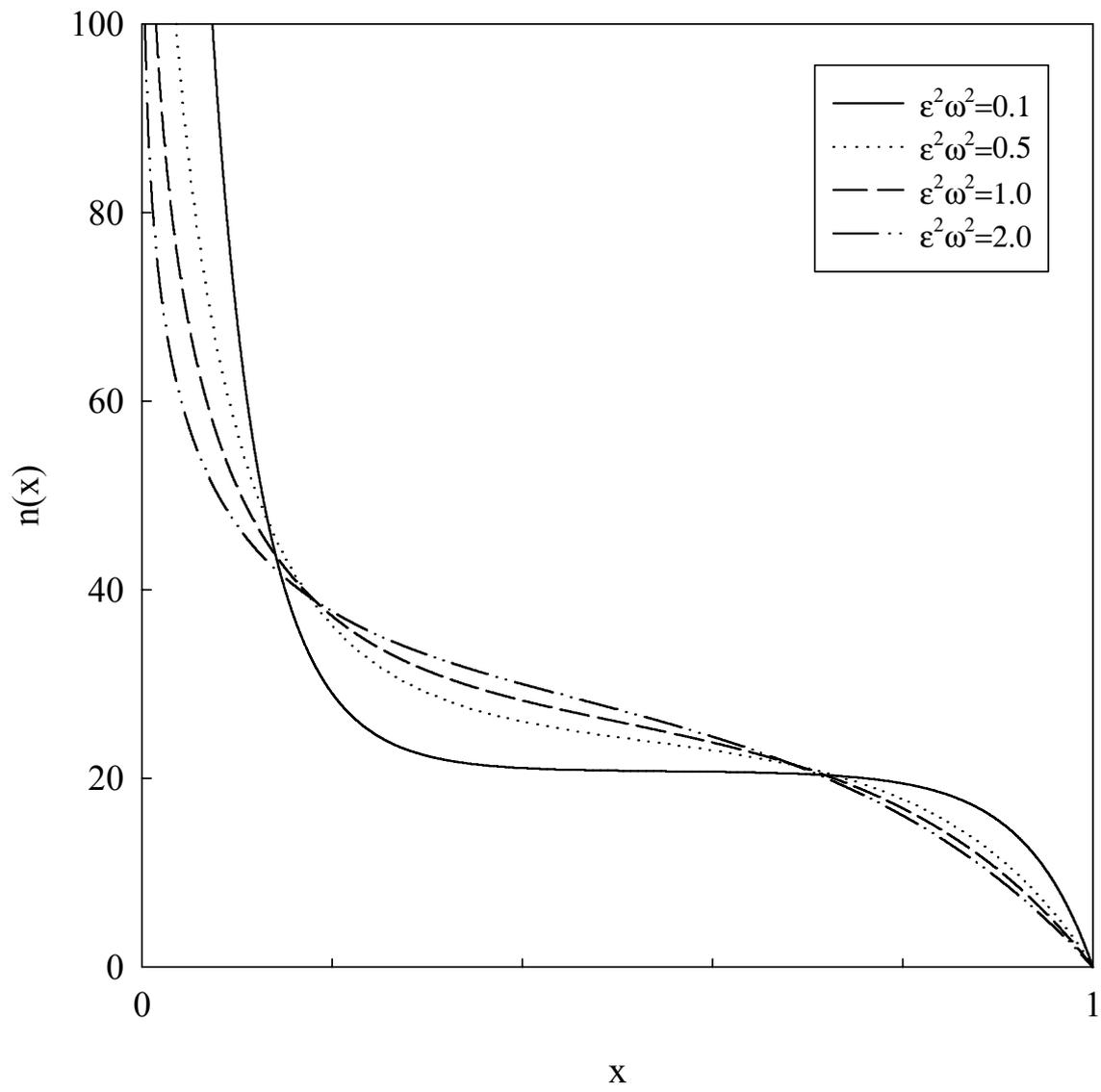

**FIG.4(a)**

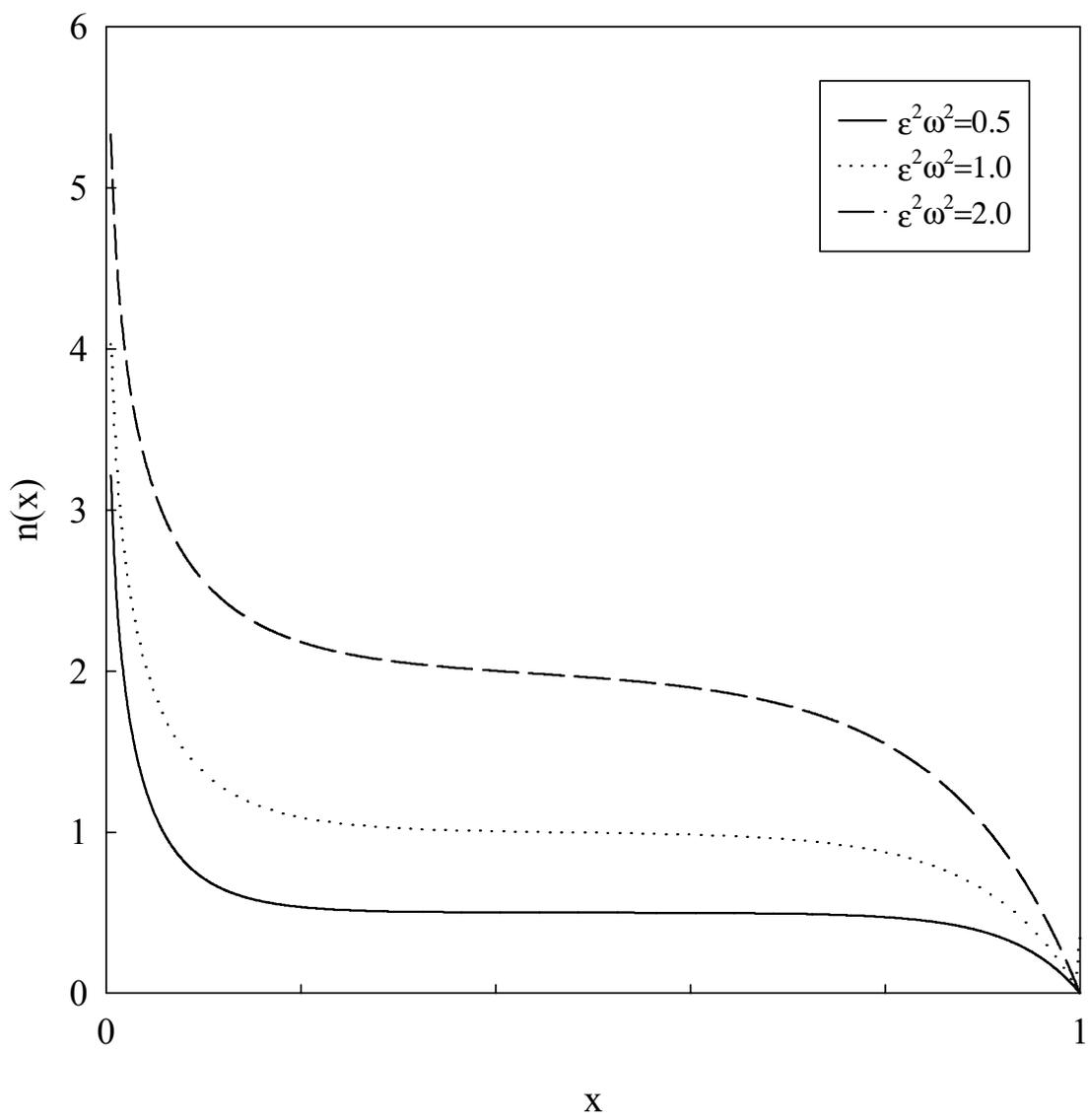

**FIG.4(b)**

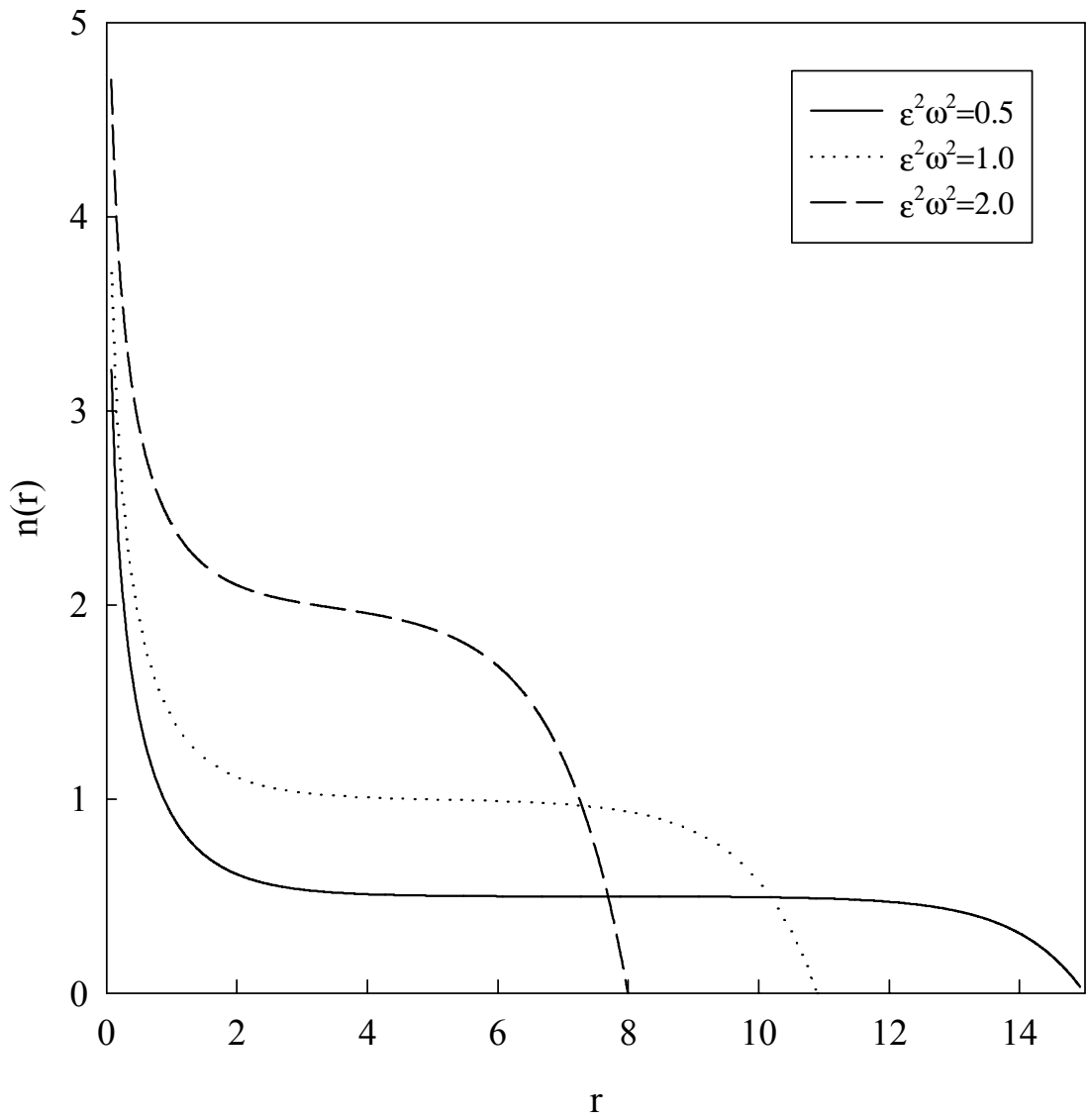

**FIG.4(c)**

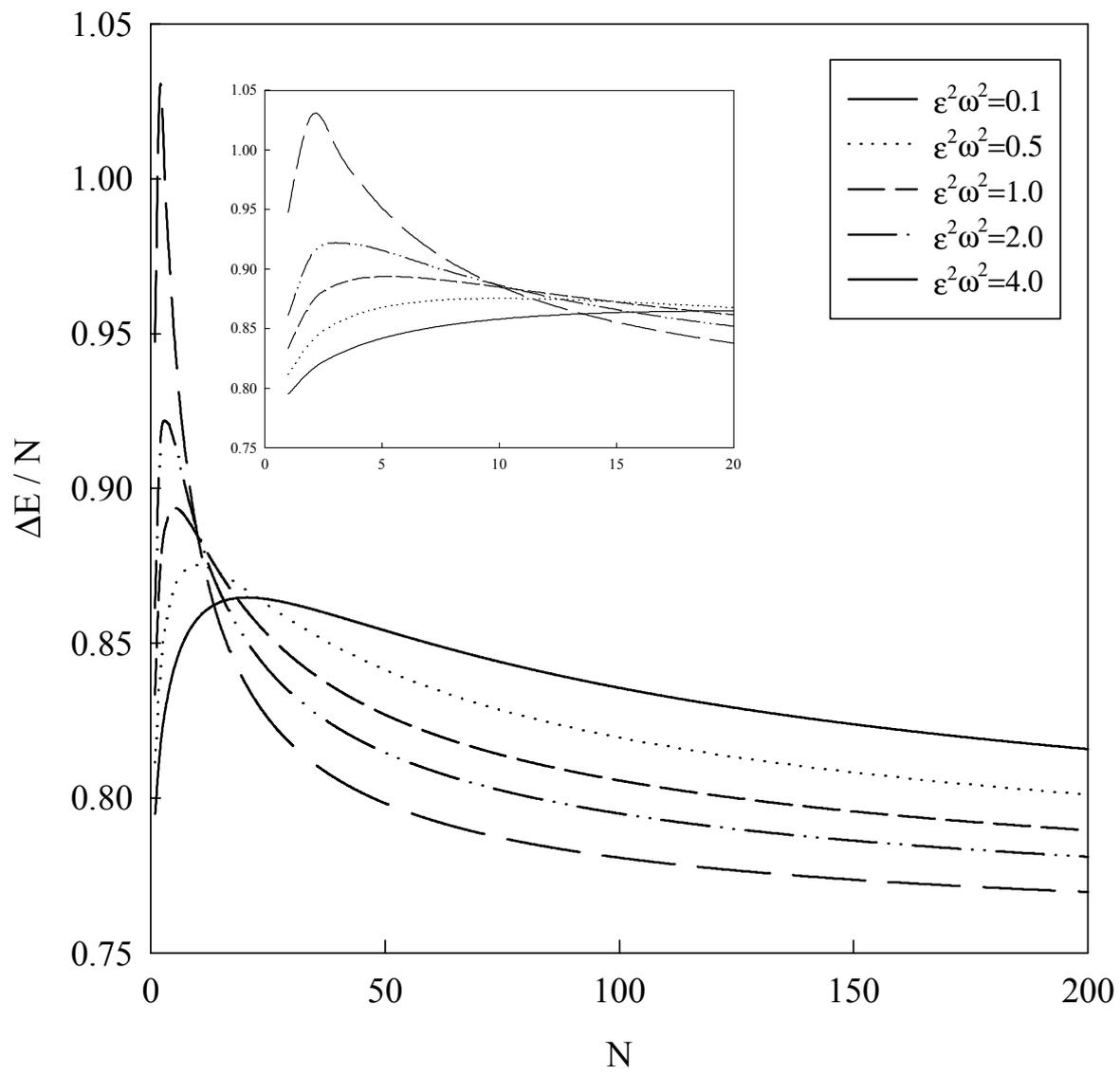

**FIG.5**